\documentclass[epj,final,numbook]{svjour}
\usepackage{amsmath,amssymb,amsbsy}
\usepackage{graphicx}
\usepackage{mcite}

\newcommand{\qp}{q_{\perp}}
\newcommand{\vqp}{\vec q_\perp}

\newcommand{\nts}{\negthickspace}
\newcommand{\mubsiguni}{\kappa q_{\perp}^4 \nts+ \nts\left(\mu\nts +
    \nts\sigma\right) {q_{x}^2}\nts+\nts\Gamma}
\begin{document}
\renewcommand{\thefootnote}{$\fnsymbol{footnote}$}
\title{The Effect of Shear Flow on the Helfrich Interaction in
  Lyotropic Lamellar Systems}  
\author{Simon W.~Marlow and Peter D.~Olmsted
}
%
%
\institute{\email{physwm@irc.leeds.ac.uk \textrm{and}
    p.d.olmsted@leeds.ac.uk}\\ Polymer IRC and Department of Physics
  \& Astronomy, University of Leeds, Leeds LS2 9JT, UK}
\date{Received: \today / Revised version: date} \abstract{ We study
  the effect of shear flow on the entropic Helfrich interaction in
  lyotropic surfactant smectic fluids. Arguing that flow induces an
  effective anisotropic surface tension in bilayers due to a
  combination of intermonolayer friction, bilayer collisions and convection, we
  calculate the reduction in fluctuations and hence the renormalised
  change in effective compression modulus and steady-state layer
  spacing. We demonstrate that non-permeable or slowly permeating
  membranes can be susceptible to a undulatory instability of the
  Helfrich-Hurault type, and speculate that such an instability could
  be one source of a transition to multilamellar vesicles.
  \PACS{ {64.70.Md}{Transitions in liquid crystals}
    \and {83.80.Qr}{Rheology of surfactant and micellar systems, associated polymers}
    \and {87.16.Dg}{Membranes, bilayers, and vesicles}
    } 
} 
\authorrunning{S.~W. Marlow and P.~D. Olmsted}
\titlerunning{The Effect of Shear Flow on the Helfrich Interaction}
 \maketitle
\section{Introduction and Overview}
\begin{figure}[htbp]
\begin{center}
  \includegraphics[width=9cm]{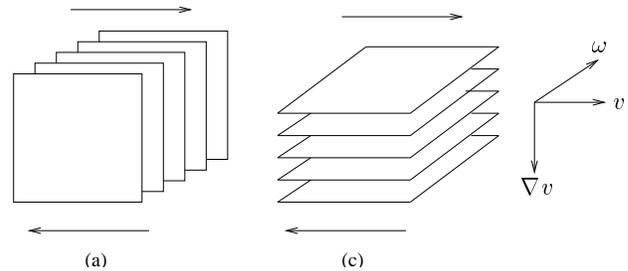}
\end{center}
  \caption{Orientations of a Lamellar Phase in Shear Flow}
  \label{fig:smectic}  
\end{figure}

Lamellar systems display a wide variety of behaviours in the presence
of shear flow. Weak flows typically align lamellae so that they slide
over one another with layer normals parallel to the flow-gradient
($\boldsymbol{\nabla}\vec{v}$) direction ($c$ orientation in Fig.
\ref{fig:smectic}).  Lamellar block-copolymer melts undergo a series
of reorientation transitions as a function of strain rate, frequency,
and amplitude, between the $c$ and $a$ (layer normals parallel to the
vorticity ($\omega$) direction) orientations
\cite{Lars+93,fredrickson94,Pate+95,Wiesner97}. Thermotropic smectic
liquid crystals also undergo a variety of reorientational transitions
as a function of strain rate \cite{safinya91,bruinsma91}. Lyotropic
smectics, such as surfactant lamellar systems, can form multi-lamellar
vesicles, or ``onions'', either above a critical strain rate (see
\cite{RND93,DRN93,PanizzaCCR98} for studies of SDS surfactant
systems), or for very small (essentially zero) strain rates (see
\cite{leon00} and \cite{BHT97} for studies of AOT and ionic systems,
respectively). At higher strain rates these onions can undergo further
transitions to onion crystals of different symmetries and sizes
\cite{SierRoux97,DRN95} as well as to oscillatory states
\cite{wunenberger01}.  Scattering experiments on some lyotropic and
thermotropic smectics (main chain and side chain) suggest other
multilamellar curved structures, such as concentric cylinders, or
``leeks'' \cite{PAR95,diat93,CristobalRCP00}. Onion formation and
layer rearrangement have been correlated with screw dislocations in
DMPC/$C_{12}E_{5}$ solutions \cite{DhezND01}. Onions have also been
seen in triblock copolymer solutions
\cite{ZipfelBSLATR99,*ZipfelLTAR99}.

The relative and absolute stability of the $a$ and $c$ orientations
has been understood from several points of view, but there is no clear
picture yet that indicates which mechanisms operate in which systems.
Cates and Milner predicted that shear can stabilise a layered system
near the equilibrium sponge-lamellar phase transition
\cite{catesmilner89}; such a transition has been reported
experimentally \cite{MMK96,YamaTana96}. Ramaswamy \cite{Ramaswamy92b}
argued that shear flow can suppress undulations in the $a$
orientation, leading to a collapse of the lamellar phase. The
dependence of the critical shear rate, $\dot\gamma_c$ on layer
spacing, $d$ and membrane viscosity $\eta$ was calculated to be
$\dot\gamma_c\sim (k_B T)^3/ \eta \kappa^2 d^3$.  Such a collapse was
seen in the flow of a thin lamellar phase of the ionic surfactant AOT
in brine, in a specially designed Couette cell assumed to shear the
layers in the $a$ orientation \cite{alkahwaji00}. The dependence on
layer spacing matched Ramaswamy's prediction, if the measured
$d$-dependence of the zero shear viscosity ($\eta\sim 1/d^2$) was
incorporated. However, viscosity measurements were performed in a
standard Couette cell rheometer; there is some evidence that AOT
orders in the $c$-orientation in such a geometry \cite{diat93a}. 
In addition a slightly different dependence of the critical shear rate
on layer spacing was found for a lamellar phase consisting of an
anionic surfactant, a deviation that might be explained by the
presence of the more dominant Helfrich interaction. To further
understand these experiments, in addition to disentangling the
viscosity issues, it may be important to consider the propagation of
the air/fluid interfacial surface tension into the bulk.

Milner and Goulian \cite{goulian95} examined the instability of
thermotropic smectics in the $c$ orientation, and noted that convection
of bending fluctuations can induce compression and destabilises
the $c$ orientation with respect to the $a$ orientation. Fredrickson
showed that non-linearities in the diblock free energy can destabilise
the $c$ orientation with respect to the $a$ orientation
\cite{fredrickson94}. Another mechanism is viscous contrast
\cite{winey93,Wiesner97}: the $c$ phase allows the strain field to be
concentrated in the less viscous material, which would be favoured
at higher frequencies or strain rates.

Although the appearance of the onion phase in shear flow is
well-documented, there is still no reliable theoretical framework for
this phenomenon. An appealing device for the onion instability is an
analogy of the Helfrich-Hurault mechanism \cite{DelayeRD73,pgdg}, in
which an applied dilational strain parallel to the layer normals may
be relieved by buckling to retain the equilibrium layer spacing.
Oswald and Kleman \cite{OswaldK82} proposed that a smectic flowing
between misaligned plates generates defects that induce the necessary
dilational strain, and Roux and co-workers used this argument to
rationalise their results \cite{RND93}.  The shear rate $\dot\gamma_c$
for buckling was estimated to vary inversely with the square of the 
gap size between
the plates. However, not only do experiments show that the gap size
does not affect $\dot\gamma_c$ \cite{diat93}, but the predicted value
for $\dot\gamma_c$ differs from the observed value by a few orders of
magnitude.  Wunenburger and co-workers \cite{WunenburgerCCR00}
extended this idea to consider the higher shear-rate transition from
onions to another well-oriented lamellar phase. They demonstrated
that, at high enough strain rates, an undulation could persist in the
velocity gradient direction while a restabilisation would occur in the
velocity direction. Reasonable comparisons were made between the
resulting theoretical stability diagrams and the experimentally
observed transitions.

In the case of thermotropic smectics, Auernhammer and co-workers
\cite{AuernhammerBP00} suggested that flow can induce a layer tilt due
to brushing of molecules in adjacent layers aligned parallel to the
normal.  This induces an effective strain due to the layer shrinkage
in the normal direction (if there are few defects and the necessary
change in layer number cannot occur) that can, in principle, be large
enough to destabilise the $c$ orientation.  Unpublished simulations
\cite{Soddemann} found undulations in the vorticity direction.

In this work we consider the entropic, collision dominated membranes first
studied by Helfrich \cite{helfrich78}, and neglect the effect
of electrostatic forces.   
Shear flow has a dramatic effect on the
fluctuation spectrum, and a reasonable proposition is that flow
suppresses fluctuations in the $c$ orientation, hence reducing
intermembrane interactions and repulsion between the layers.  For
systems with many defects, this reduced repulsion would be expected to
lead, after an appropriate time, to a reduction in mean layer spacing.
Indeed not only was this seen by Yamamoto and Tanaka
\cite{YamaTana95b}, but also the shape of the x-ray scattering
Bragg peak became broader and less intense.  For systems with few
defects or very slow permeation, the reduced fluctuation spectrum
would lead to an effective imposed strain along the layer normals,
which if sufficiently large could be relieved by buckling.  This,
then, could be a mechanism leading to onion formation, in the event
that other considerations allow for onions. For example, the Gaussian
curvature modulus $\bar{\kappa}$ should not be too positive, which
would tend to suppress structures with positive mean curvature, and
the $a$ orientation should also be dynamically unstable.

Our task, then, is to calculate the suppression of layer undulations
in the $c$ orientation, and hence the renormalised ``interlayer
potential'' in steady state shear flow. In lieu of potentials, the
proper starting point is the full dynamics of a lyotropic smectic
\cite{nallet90,*Nall+94,*prostmann98} in the presence of an imposed
boundary stress or strain rate.
Rather than taking this
approach, we assume a coarse-grained free energy, following Bruinsma
and Rabin \cite{bruinsmarabin92}, as a first step towards
understanding the effect of flow.  In common with the recent work of
Zilman and Granek \cite{zilman99}, we assume that bilayers oriented in
the $c$ orientation experience a flow-induced effective tension,
although the tension that we propose is anisotropic and
has a different sign and physical interpretation (Appendix \ref{sec:ZG work}).
In Section~\ref{sec:tension} we explore the effects of this tension on
the steady-state layer spacing $d$ and the compression modulus
$\bar{B}$ of an equilibrium Helfrich stabilised lamellar phase. 
In Section~\ref{sec:onions} we examine these results in the context of
an imposed shear flow. After estimating possible sources of tension 
we discuss the predicted change in layer spacing and the possibility 
for an instability, taking into account permeation and defect-creation. 
We finish with a summary in Section~\ref{sec:conclusion} and outline
future work that will quantitatively address the dynamic origin and 
consequences of this tension. 

There have been several discussions of the effect of flow on the layer
spacing in block copolymer lamellar systems, which focus on the change
in polymer conformations due to the stretching imposed by flow in
melts and solutions \cite{williamsmack94,*harden96,*grest99}, and layer
changes have been reported in melts \cite{Poli+99}.  Our mechanism is
unrelated to this one, but may be related to recent experiments on
block copolymer solutions that have demonstrated a change in layer
spacing \cite{PopleHD98} and onion formation as a function of flow
\cite{ZipfelBSLATR99}, if those systems are Helfrich stabilised.
\section{The Lamellar Phase under Tension}\label{sec:tension}
\subsection{The Equilibrium zero-tension Lamellar Phase}\label{sec:eqm}
The equilibrium free energy $F$ of a three-dimensional lyotropic
smectic fluctuating at fixed chemical potential penalises gradients in
the mean layer displacement $u(\vec{r})$:
\begin{equation}
   F = \tfrac12\int\!d^3\!r \left[K(\nabla^2_{\perp}u)^2 + \bar{B}(\partial_z
      u)^2\right].  \label{eq:5}
\end{equation}
The elastic constants in Eq.~(\ref{eq:5}) are the
bending modulus $K$ and the compression modulus at fixed chemical
potential $\bar B$. For membranes that interact via collisions the
origin of $\bar B$ is the steric repulsive interaction
\cite{helfrich78}; the collisions induce an 
 entropic confinement pressure $p$, 
\begin{equation}
  \label{eq:6}
  p \sim \frac{k_B T}{L_p^2 d},
\end{equation}
where $L_p$ is the characteristic distance between collisions (the
``patch'' or ``collision'' length) and $d$ is the layer spacing, from
which the compression modulus may be estimated by
$\bar{B}\sim -d\partial p/\partial d$

The patch length is the
implicit lower cut-off of the integral in Eq.~(\ref{eq:5}) and is that
transverse distance over which membranes wander before colliding,
found by computing the mean fluctuations of a free membrane,
governed by the single membrane free energy $f$. In the Monge gauge,
\begin{eqnarray}
  \label{eq:7}
  f = {\tfrac12}\kappa \int_{A_{\perp}} d^2 \!r  \left(\nabla^2_{\perp} h({\bf
    r})\right)^2 
   = {\tfrac12}\kappa L^2 \sum_{\vqp} {\left| h_{\vqp} \right|}^2 q_{\perp}^4, 
\end{eqnarray}
where $h(x,y)=h(\bf r)$ is local height of a given membrane, rather
than the mean position $u(\vec{r})$ that we introduced above for the
coarse-grained three-dimensional free energy. $A_{\perp}$ is the
projected membrane area orthogonal to the mean membrane normal.
$L$ is the system size and we have defined the Fourier Transform $h(\vec
r)=\sum_{\vqp} \!h_{\vqp}e^{i{\vec {{\vqp}\cdot r}}}$.  Assuming periodic
boundary conditions,
\begin{equation}
  \label{eq:7c}
  \sum_{\vec q_{\perp}}  = {\left(\frac{L}{2\pi}\right)^2}
  \int_{\pi/L_p}^{\pi/a} d^2{\vec q_{\perp}},
\end{equation}
where $a$ is a molecular size and wavelengths longer than $L_p$ are
suppressed by steric hindrance with neighbouring membranes. Note that
\begin{equation}
  \label{eq:22}
  u(x,y,z=nd)=\int_{A(L_p)} \left[h(x'-x,y'-y) -
  nd\right]\,\frac{d^2r'}{A(L_p)} 
\end{equation}
is the average of the microscopic layer displacement over a patch area
$A(L_p)$ centred at $(x,y,z=nd)$, so that Eq.~(\ref{eq:5}) encompasses
Eq.~(\ref{eq:7}) subject to a hard wall constraint and $K={\kappa/
  d}$.

We neglect the Gaussian curvature, penalised by the modulus
$\bar{\kappa}$, which is valid for fixed topology.  However,
transitions to a different topology such as onions, or changes in the
 distribution of dislocations or pores \cite{MinewakiKYI99}, may
involve $\bar\kappa$. Quantities known to alter $\bar{\kappa}$, such
as salinity, temperature, or cosurfactant, seem to influence the shear
rate and/or the mechanism for onion formation
\cite{leon00,ZipfelNLLOR01,LeOMZR01,ZipfelBLR99}.

The compression modulus may be estimated by relating $L_p$ to $d$
\cite{helfrich78}. The constraints of neighbouring membranes
limits the mean square fluctuations to the mean layer spacing $d^2$,
\begin{equation}
  \label{eq:7e}
\langle h^2(\vec{r})\rangle =\alpha d^2,
\end{equation}
where $\alpha$ is a constant of proportionality. The calculation of
the mean square fluctuations is straightforward:
\begin{equation}
  \label{eq:7bb}
  \langle h^2(\vec{r})\rangle
   =\sum_{\vqp}\langle {\left| h_{\vqp} \right|}^2\rangle 
   =\frac{k_B T}{4\pi^3\kappa} L_p^2,
\end{equation}
where $L_p/a \gg 1$.  The constraint on the height fluctuations
Eq.~(\ref{eq:7e}) leads to an expression for the collision length,
\begin{equation}
  \label{eq:7f}
  L_{p}=cd\sqrt\frac{\kappa}{k_B T} .
\end{equation}
If $\alpha = 1/{3 \pi^2}$, as inferred by Helfrich \cite{helfrich78}, 
then $c=\sqrt{4 \pi / 3}$.
There is some debate about the exact value of $c$; for example,
Golubovic and Lubensky \cite{GolubovicL89} calculated
$c=\sqrt{32/3\pi}$. 

Combining the collision length with the 
pressure (Eq.~\ref{eq:6}) yields a compression modulus that has been
calculated precisely as \cite{helfrich78,BachmannKP01,JankeKM89} 
\begin{equation}
  \label{eq:7g}
  \bar{B}= \frac{6\delta_n n}{n+1}\frac{(k_B T)^2}{\kappa d^3},
\end{equation}
where $n$ is the number of membranes. For $n=\infty$ Helfrich
determined $\delta_\infty = 3\pi^2/128\sim 0.23$ \cite{helfrich78}.
Strong coupling perturbation calculations \cite{BachmannKP01} suggest
that $\delta_\infty\sim 0.1$ and for a single membrane,
$\delta_1=\pi^2/128\sim 0.07$; both results are consistent with those
given by Monte-Carlo simulations \cite{JankeKM89}.

For flat membranes in a lamellar phase the volume fraction is given by
$\phi=t/d$, where $t$ is the layer thickness. For fluctuating lamellar
phases of uniform thickness $t$, the mean volume fraction is
\begin{equation}
  \label{eq:9}
  \phi = \left\langle\frac{A}{d}\right\rangle\frac{t}{A_{\perp}},
\end{equation}
where $A$ is total membrane area. To find the relation between layer
spacing and concentration, we can expand (for given mean spacing $d$)
$A\simeq A_{\perp}(1+\tfrac12\langle(\nabla_{\perp}h)^2\rangle$, to
find
\begin{subequations}
    \label{eq:9d}
\begin{align}
  d&=\tfrac{t}{\phi}\left[1 +
    \tfrac12\langle(\nabla_{\perp}h)^2\rangle\right] \label{eq:20}\\
&=\frac t \phi \left[1+{k_B T\over {4\pi \kappa}}{\ln
        \left(c{d\over a}\sqrt{\frac {\kappa} {k_B
              T}}\right)}\right]. \label{eq:23} 
\end{align} 
\end{subequations}
Membranes dominated by electrostatic interactions are essentially
flat, in which case $d\phi$ is independent of the volume fraction
$\phi$.  Hence, deviation from this `ideal' dilution law according to
Eq.~(\ref{eq:9d}) suggests that the membranes are stabilised by the
Helfrich interaction \cite{RouxNFPBSM92}.  Use of Eq.~(\ref{eq:9d}),
given $d$ and $t$, allows $\kappa$ and $a$ to be extracted from
swelling experiments.  Helfrich systems include various dilute
non-ionic and screened surfactant solutions. Whether or not certain
dilute copolymer solutions that form onions under shear
\cite{ZipfelBSLATR99} are stabilised by the Helfrich interaction is
unclear.  Further evidence of Helfrich stabilisation is deduced from
the power law behaviour exhibited by the x-ray structure factor
\cite{rouxsafinya88}, from which the bending modulus can be extracted.
\subsection{Lamellar Phase subjected to  a Finite Anisotropic Tension}
\subsubsection{Equilibrium Layer Spacing} \label{sec:dtension}
When a Helfrich-stabilised lamellar phase is sheared in the $c$
orientation, layers are convected and stretched by the flow, adjacent
fluctuating layers collide, and flow induces the leaves of the bilayer
to slide over one another. 
A rigorous study of the effect of flow on membranes should consider 
appropriate equations of motion and treat the
dynamics explicitly. 
For example, the best approach, albeit probably intractable,
would be to coarse-grain a dynamical description of coupled monolayers
such as that of Ref.~\cite{SeifLang94,*SeifLang93} to an effective 
two-component smectic theory, incorporating both flow and
the normal and tangential collision forces. 
We adopt instead a `quasi-equilibrium` approach 
to gain insight into the stability of the steady states for membranes in
flow. We propose that the resulting interplay of
hydrodynamic interactions and interbilayer friction can be described
by a flow-induced effective tension acting parallel to the layers.

In the presence of a tension, Seifert's \cite{Seif95} self-consistent
calculation of the potential between a wall and a membrane of a
vesicle under tension might be applied to find the preferred
separation.  However, since repulsive interactions from neighbouring
membranes dominate the interlamellar potential far from the unbinding
transition \cite{milnerroux92}, a similar derivation of the layer
spacing cannot be achieved. Thus we are led to deduce a new layer
spacing without regard for the change in potential by generalising
Helfrich's geometric argument leading to the dilution law
Eq.~(\ref{eq:9d}) to include a lateral tension \cite{helfrich84}.

To retain the asymmetry between flow and vorticity directions in the plane,
we model the flow as an effective anisotropic
tension applied in the flow direction.
The free energy of a membrane subject to an anisotropic tension
$\sigma$ is
\begin{align}
  \label{eq:12a}
  f&= {{1\over 2}\int d^2r \left[\kappa({{\nabla}^{2}_{\perp} h})^2
      + \sigma {(\boldsymbol{\nabla}_{x}h)}^2 \right]} ,
\end{align}
where $\sigma$ penalises short wavelength modes and the increased area due to
stretching the membrane in the $x$-direction. This decreases the excess
area and suppresses fluctuations:
\begin{subequations}
 \begin{align} 
 \label{eq:12b}  
  \langle h^2(\vec{r})\rangle & =
  {k_BT \over (2\pi)^2}\int_{\pi/L_p}^{\pi/a}
  \frac{d^2 \vec{q}_{\perp}}{{\kappa q_{\perp}^4}+
  {\sigma q_x^2}}\\ 
  &={k_B T \over {2\pi\sigma}}\left(\sqrt{1+\frac{\sigma L_{p}^2}
  {\kappa\pi^2}}-1\right),\label{eq:12bb}
 \end{align}
\end{subequations}
 for $L_p/a \gg 1$. 
 Here we assume an isotropic change in patch length, rather than
 the anisotropic renormalisation that would be expected; this should 
 suffice for qualitative behaviour, as we show below. Assuming the
 same hard wall constraint as without tension, Eq.~(\ref{eq:7e}), we
 can calculate a renormalised mean patch length,
\begin{equation}
  \label{eq:12c}
   L_{p}^2 =4 \pi^3 \alpha d^2 \frac{\kappa}{k_B T}\left(1+
   \alpha\pi\frac{\sigma d^2}{k_B T}\right).
\end{equation}

By the equipartition theorem,
\begin{subequations}
 \begin{align}
  \label{eq:12d}
  \langle (\boldsymbol{\nabla}_{\perp} h(\vec{r}))^2\rangle =
  {k_BT\over {(2\pi)^2}}\int_{\pi/L_p}^{\pi/a}
  \frac{q_{\perp}^2 d^2 \vqp }{{\kappa q_{\perp}^4}+
  {\sigma q_{x}^2}}\\  
  ={k_B T \over {2\pi\kappa}}
   \left[\ln\frac{L_p}{a}-
   \ln{\left|\frac{1+\sqrt{1+\frac{\sigma L_{p}^{2}}{\kappa\pi^2}}}
   {1+\sqrt{1+\frac{\sigma a^2}{\kappa\pi^2}}}\right|}\right].\label{eq:12dd}  
 \end{align}
\end{subequations}
The simultaneous conditions relating the patch size to the layer
spacing via $\langle h^2\rangle$ (Eq.~\ref{eq:12b}), and the
concentration to the degree of crumpling $\langle
(\nabla_{\perp}h)^2\rangle$ (Eq.~\ref{eq:20}) determine the layer
spacing as a function of concentration and tension,
$d(\phi,\sigma)$, as well as the associated collision length
$L_p(\phi,\sigma)$. Hence, using Eq.~(\ref{eq:12c}) to eliminate
$L_p$\footnote{For an isotropic tension the swelling law is
  $\hat{d}=\hat{d}_0+{1\over {8\pi k}} \left\{\ln\left[1+\frac{\hat
        \sigma}{\hat d_0^2}\frac{1}{k} \left(\frac{\phi}{c \hat
          t}\right)^2\right] - \ln\left[\hat \sigma + \hat\sigma
      \left(e^{\hat{\sigma}\hat{d}^2/\hat{d}_0^2} -
        1\right)^{-1}\right]\right\} .$ }, we find
\begin{equation}
  \label{eq:12e}
  \hat{d} = \hat{d}_0 + {1\over {4\pi k}} 
  \ln{\left|\frac{\hat{d}}{\hat{d}_0} 
  \left(\frac{1+\sqrt{1+\frac{\hat\sigma}{\hat{d}_0^2}\frac{1}{k}
  \left(\frac{\phi}{c \hat{t}}\right)^2}}
  {2\sqrt{1+\frac{\hat{\sigma}}{4}
  \left(\frac{\hat{d}}{\hat{d}_0}\right)^2}} \right)\right|} 
\end{equation}
where $\hat{d}={d\phi/ t}$, $\hat{t}={t/ a}$, $k={\kappa/ k_B T}$ and
$\hat{\sigma}=4\pi\alpha \sigma d_0^2/k_B T$ are non-dimensionalised
parameters and $d_0$ is the layer spacing at zero tension.  As
expected, an increasing tension reduces the excess area.

\begin{figure}[!htbp]
   \includegraphics[width=8.5cm]{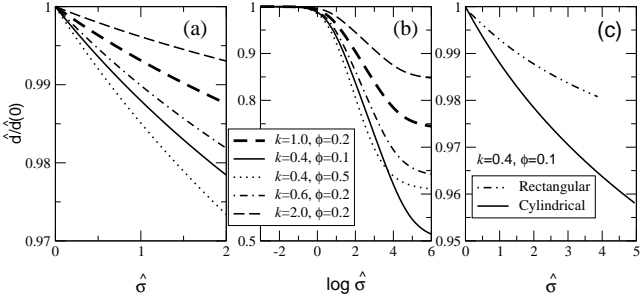}
      \caption{Fractional change in layer spacing $\hat{d}/
        \hat{d}_0$ for the isotropic (cylindrical) patch length
        renormalisation, from Eq.~(\ref{eq:12e}) and Eq.~(\ref{eq:9d})
        at $\hat{t}=6$ as a function of induced tension $\hat\sigma$
        for small values (a) and $\log\hat\sigma$ for all values (b).
        (c) A comparison of cylindrical (Eq.~\ref{eq:12e}) and
        rectangular (Eqs.~\ref{eq:25}) patch size renormalisations,
        for $k=0.4,\, \phi=0.1$.  Note $k=\kappa/k_B T$ and
        $c=\sqrt{4\pi/3}$. }
      \label{fig:1}
\end{figure}

For $\hat{\sigma}\ll1$ we can calculate the change in the preferred
layer spacing,
\begin{align}
  d & = d_0 \left[1-\frac{\sigma d_0^2}{k_BT}
    \left(\frac{\frac12-\frac{k_B T}{\kappa}\left(\frac{a}{c d_0}\right)^2 }
      {\frac{4\pi\kappa}{k_BT}\frac{\phi}{t} d_0 -1}\right)
    \pi\alpha+ \ldots\right].
  \label{eq:12f} 
  \end{align}
Note that the linear correction is half that of a membrane subjected
to isotropic tension. The asymptotic expression for high tension is
\begin{equation}
  \label{eq:12g}
  d=\frac{t}{\phi}\left(1+{k_B T\over 4\pi \kappa}
  \frac{c}{a}\sqrt{\frac{\kappa}{4\pi\alpha\sigma}}\right).
\end{equation}
The relative reduction in layer spacing, calculated from
Eq.~(\ref{eq:12e}), is shown in Fig.~\ref{fig:1}.  This applies
directly to equilibrium systems with a mechanically-applied tension,
such as few-layered vesicles or bilayers tethered to surfaces.
Tension should have the greatest effect when acting on the largest
excess surface area, and hence we expect a greater decrease in layer
spacing for a lower bending modulus and large crumpling fraction.

As mentioned above, we have assumed that the patch length increases
isotropically with increasing tension, and hence imposed a cylindrical
geometry with a change in the lower radius cutoff $\pi/L_p$.
Physically, one would expect, at least for small tensions, that the
patch size would actually increase anisotropically, with fluctuations
in the transverse direction less affected by stretching.  In this case
one should impose a rectangular domain of integration, with a cutoff
$\pi/L_{py}$ for the $q_y$ integration, which remains fixed for
increasing $\sigma$, and a cutoff
$\pi/L_p(\phi,\sigma)\leq\pi/L_{py}$ for the $q_x$ integration,
which decreases for increasing $\sigma$. In this case the
simultaneous set of equations that determines the new layer spacing
$d$ and patch size $L_p$ is:
\begin{subequations}
\label{eq:25}
\begin{align}
  \alpha d^2 &= \displaystyle
\frac{k_BT}{4\pi^2}\!\!\!\!\mathop{\int}_{\pi/L_{py}}^{\pi/a}\!\!\! dq_y 
    \!\!\! \!\!\!\mathop{\int}_{\pi/L_p(\phi,\sigma)}^{\pi/a}\!\!\!
    \!\!\! \!\!\!dq_x\,  
      \frac{1}{\kappa (q_x^2 + q_y^2)^2 + 
        \sigma q_x^2}\\
\frac{\phi d}{t} &= \displaystyle 1 + \frac{k_BT}{8\pi^2}\!
\!\!\!\mathop{\int}_{\pi/L_{py}}^{\pi/a}\!\!\!dq_y 
    \!\!\! \!\!\!\mathop{\int}_{\pi/L_p(\phi,\sigma)}^{\pi/a}\!\!\!
    \!\!\! \!\!\!dq_x\,  
      \frac{(q_x^2 + q_y^2)}{\kappa (q_x^2 + q_y^2)^2+
        \sigma q_x^2}
\end{align}
\end{subequations}
Results of this calculation for a representative parameter set 
are shown in Fig.~\ref{fig:1}c. As
expected, the inclusion of an anisotropic patch size leads to a
smaller layer shrinkage, because the transverse fluctuations remain
substantial. Note, however, that solutions do not exist above for 
$\hat{\sigma}\simeq4.0.$
Formally, this limit corresponds to $L_p\rightarrow\infty$ with $L_y$
finite, and describes layers that are essentially flat in the tension
direction while crumpled in the transverse direction. Microscopically,
membranes are essentially incompressible, so this limit is unphysical
and we expect the layer contraction to evolve towards the isotropic
patch calculation with increasing stress.
\section{Application to Membranes in Flow}\label{sec:onions}
\subsection{Physical Picture}
\begin{figure}[htbp]
  \centerline{\includegraphics[width=8cm]{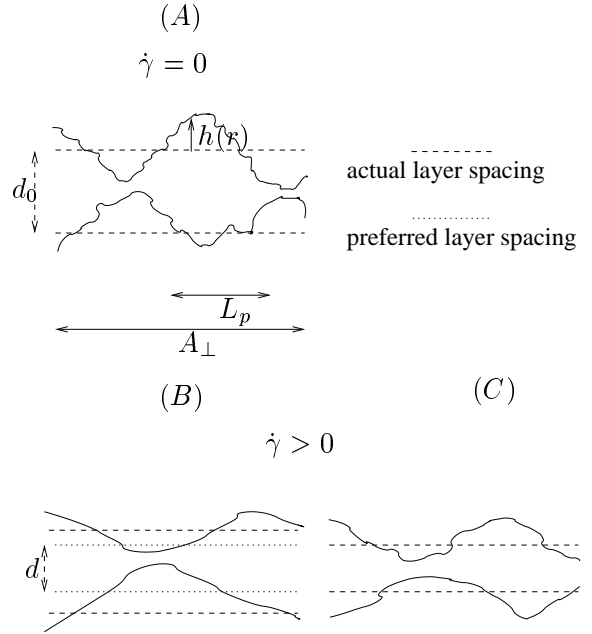}}
\caption{Two
  different scenarios for a lamellar phase (A) subjected to flow. In
  (C) fluctuations are reduced and the layer spacing reduces due to
  the softened repulsive potential, while in (B) the equilibrium layer
  spacing is maintained because the system cannot relax (and form more
  layers).}
\label{layers}
\end{figure}

We consider a Helfrich-stabilised lamellar phase with layer spacing
$d_0$ (Fig.~\ref{layers}(A)). In flow, transverse membrane
fluctuations are suppressed, leading to fewer collisions and hence
more ``space'' between layers.  If there is no permeation or other
mechanism to change the layer number on the experimental time scale,
the layer spacing remains fixed at its initial value
(Fig.~\ref{layers}(B)). At sufficiently large strains (Section
\ref{sec:instability}) an instability will relieve the strain in
favour of undulations via a non-equilibrium version of the
Helfrich-Hurault effect.  This then could be a precursor to either a
stable modulated phase or an instability to onion formation.  However,
significant permeation due to passages, trans-membrane diffusion, or
smectic defects, can lead to a layer spacing reduction after
sufficient time for the required increase of layer number, as in
Fig.~\ref{layers}(C). Which effect is seen depends on the rate of
application of flow, the permeability and the defect structure, and the
degree of crumpling.
\subsection{Sources of Tension}\label{sec:sources}
Under the assumption that flow induces an effective tension in a
lamellar phase, we apply the results of Section~\ref{sec:dtension}.
This is heuristic, but we believe yields qualitatively useful results,
as employed by Zilman and Granek \cite{zilman99}.  Dynamically,
tension corresponds to a tangential force resisting changes in total
membrane area that induces a local normal force on the membrane.
Hence, we contend that shear flow ``irons out'' small scale membrane
wrinkles.  Several dynamic effects can lead to an effective tension:
(i) shear flow distorts and stretches membrane fluctuations, (ii)
friction between the leaves of the bilayers increases the drag due to
intra-bilayer dissipation in undulations, and (iii) intermembrane
collisions induce a transverse membrane force. The first two effects
are non-linear and yield tensions that in the simplest case scales as
$\dot{\gamma}^2$, while the latter effect is linear in $\dot{\gamma}$.
All of these mechanisms are expected to be unimportant for flat,
non-Helfrich membranes with weak equilibrium undulations.

The importance of convection and stretching can be estimated by
balancing the restoring force on a wrinkle of wavenumber $q$ with the
viscous dissipation encountered in the flow in a thin film to find the
typical undulation lifetime \cite{bruinsmarabin92}. Fluctuations are
influenced when the strain rate is of order the inverse lifetime of
the slowest relevant undulations, which are those of wavelength the
patch size $L_p$. This leads to an estimate for the critical strain
rate $\dot{\gamma}_c$ at which undulations are appreciably suppressed
by flow \cite{bruinsmarabin92},
\begin{equation}
  \label{eq:1}
  \dot{\gamma}_c\sim \frac{(k_B T)^{5/2}}{\eta d^3\kappa^{3/2}}
\end{equation}
where $\eta$ is the solvent viscosity.

A more careful calculation begins with, for example, the Langevin
equation for the affine convection of a fluctuating
membrane\footnote{In the dynamical description of the long wavelength
  variable $u(\vec{r})$  the convective term is $\dot{\gamma} z
  \partial_x u$, appropriate for wavelengths larger than the layer
  spacing and for fluctuation amplitudes $u$ smaller than the layer
  spacing \cite{goulian95,bruinsmarabin92}.} in the $c$ orientation in
a shear field $\vec{v}=\dot{\gamma} z \hat{\vec x}$,

\begin{equation}
  \label{eq:2}
  \begin{split}
  \left[\partial_t+
    \dot{\gamma}h\left(\vec{r},t\right)\frac{\partial}{\partial
    x}\right]&\,h(\vec{r},t) =\\ 
  -\kappa\int\,d^2r'\Gamma &\left(\vec{r}-\vec{r}'\right)
  \nabla^4h(\vec{r}',t) + \xi(\vec{r},t).
\end{split}
\end{equation}
The kinetic coefficient $\Gamma \left(\vec{r}-\vec{r}'\right)$ depends
on the details of the fluid-membrane coupling, with the form (in
Fourier space) $\Gamma_{\qp}\sim \eta^{-1} l^{\nu+1} q^{\nu}_{\perp}$.
Here we give three examples of relaxation mechanisms that alter the
exponent $\nu$ and the associated length scale $l$:
  \begin{equation}
    \label{eq:12}
    \Gamma_{\qp} =
   \begin{cases}
     \eta^{-1}\zeta \qp^{0}& \text{permeable}\\
     \eta^{-1} q^{-1}_{\perp}& \text{isolated}\\
     \eta^{-1} d^3 q^2_{\perp}& \text{confined fluid/squeezing}.\\
   \end{cases}
  \end{equation}
  The permeation length scale $\zeta$ depends on the size and the
  density of the pores and the membrane thickness\footnote{A simple
    model of flow through circular pores of width $w$ leads to
    $\zeta\sim t\psi^2/w^4$, where $\psi$ is the mean pore separation
    within the membranes.}.  For wavevectors $\qp\zeta\ll 1 \ll \qp
  L_p$ the membrane may be considered impermeable.  In the confined
  fluid regime ($\qp d\gg 1$), squeezing of solvent within the
  confines of the surrounding membranes leads to $\nu=2$
  \cite{Broch-Lenn75}. The case of $\nu=-1$ describes the hydrodynamic
  interaction of an isolated membrane in solvent ($\qp d \ll 1$)
  \cite{MessagerBP90}, with a length scale fixed by the wavevector.
  For permeable membranes, the characteristic length $\zeta< d$, with
  relevant wavevectors $\qp\epsilon \ll 1 \ll \qp\zeta$, where
  $\epsilon$ is some small length scale at which this dynamical
  description breaks down.
  
  The noise $\xi(\vec{r},t)$ may or may not be related to
  $\Gamma(\vec{r}-\vec{r}')$ through the fluctuation-dissipation
  theorem, depending on the strength of the applied shear.
  Eq.~(\ref{eq:2}) is a variation of an anisotropic Burger's equation
  with a convective non-linearity \cite{bray01a,*bray01b}.  Upon
  coarse-graining Eq.~(\ref{eq:2}) up to the collision length $L_p$,
  the convective non-linearity generates a dynamic response analogous
  to an effective scale-dependent tension,
\begin{multline}
  \label{eq:13}
  \partial_t h_{\vec{q}_{\perp}}
    +i\dot{\gamma}\sum_{\vec{k}_{\perp}}(q_x - k_x)
    h_{\vec{k}_{\perp}}h_{\vec{q}_{\perp}-\vec{k}_{\perp}}  
    =\\ -\left[\Gamma_{\qp}\!\!\kappa q^4_{\perp} 
  + \Gamma_x(L_p ,\mu) \dot{\gamma}^{\mu}q_x^2 \right]h_{\vqp},
\end{multline}
where $\Gamma_x(L_p , \mu)$ depends on the particular relaxation
mechanism. Thus,
\begin{equation}
  \label{eq:19}
  \sigma_{\mathit{conv}} 
  \sim \frac{\Gamma_x(L_p , \mu)}{\Gamma_{\qp}}\dot{\gamma}^{\mu}.
\end{equation}
For example, for isolated membranes we find a ``tension'' of order
\begin{equation} 
  \label{eq:3} 
  \sigma_{\mathit{conv}} \sim 
   \frac{k_B T}{\kappa^2}L_p^5\eta^2\dot{\gamma}^2\qp .
\end{equation} 
This calculation, along with results for the other relaxation
mechanisms will be presented elsewhere \cite{marlowunpub}. A similar
quadratic scaling was also estimated by Zilman and Granek
\cite{zilman99}, based on energetic arguments. It is important to
recognize that, although the ``tension'' above applies, strictly, only
to wavelengths of order the collision length, it is generated at all
wavelengths larger than the smallest cutoff and grows during the
coarse-graining procedure. We have chosen to model this tension, which
is wavenumber-dependent, as an average value that applies for all
wavenumbers. This certainly changes any quantitative predictions, but
does not influence the qualitative aspects of our results.

Comparison of the bending and tension relaxation mechanisms of 
Eq.~(\ref{eq:13}) leads to an estimate
similar to Eq.~(\ref{eq:1}) for
the strain rate $\dot\gamma_c$ at which fluctuations are significantly
suppressed 
  \begin{equation}  
    \label{eq:10a} 
   \dot\gamma_c \sim \frac{k_B T}{\eta d^3}. 
  \end{equation}

If we consider a more detailed picture of the bilayers there are at
least two more conceivable sources of tension. The two leaves of the
bilayer exert transverse intermonolayer forces as they slide over
each other. For flat membranes this simply renormalises the solvent
viscosity; for highly crumpled membranes the viscosity is
significantly enhanced,
\begin{equation}
  \label{eq:4}
  \eta_{\mathit{eff}} \simeq \eta \left[1 + \frac{t\,\eta_b}{d\,\eta}\langle
    \left(\boldsymbol{\nabla}_{\perp} h\right)^2\rangle\right],
\end{equation}
where $\eta_b$ is the bulk viscosity of the bilayer material
(typically alkyl tails). In this highly crumpled case there
may also be an effective tension, or restoring force against the 
dissipation induced by bilayers with enhanced crumpling. 

A third contribution comes from the tangential force incurred in
interbilayer collisions. Because the bilayers retain their integrity
they are moving at the velocity of the mean bilayer position, rather
than at the velocity that would be determined by affine flow. The
enhanced local shear rate near the collision event, over a length of 
order a bilayer thickness rather than the interlayer spacing, leads 
to an excess tangential force per unit length due to collisions, 
or equivalently a tension,
\begin{equation}
  \sigma_{b-b}\sim \eta \frac{(d-t)}{t}\dot{\gamma}
  \frac{A_{\mathit{con}}}{L_p}.  \label{eq:14} 
\end{equation}
Here, $A_{\mathit{con}}$ is the collision contact area, expected to be
quite small. 
\subsection{Scenarios in Flow}\label{sec:scenarios}
\subsubsection{Layer Spacing Change}\label{sec:spacingchange}
As mentioned in Section \ref{sec:dtension}, an effective tension is
expected to change the mean layer spacing in smectics with sufficient
permeability.  Specifically, either dislocation loops must be easily
generated or the \textit{lipid} permeability must be high enough for
new layers to form before any incubation time for an instability or
other transformation, such as to onions.  Note that simple solvent
permeability is not enough. Hence, we expect that highly defective
membranes are most likely to exhibit a change in layer spacing under
flow. An example of such a system is $C_{12}E_5$, which is known to be
highly permeable in regions of the phase diagram \cite{MinewakiKYI99}
and dominated by the Helfrich interaction.  In this case one expects a
change in layer spacing governed by Eq.~(\ref{eq:12f}), with the
tension $\sigma$ given by a combination of collisions
(Eq.~\ref{eq:14}) and convective effects (Eq. \ref{eq:3}),
schematically of the form
\begin{equation}
  \label{eq:8}
  \sigma = \eta\dot{\gamma}\left[A + B\dot{\gamma}^{\mu -1}\right],
\end{equation}
where $A$ and $B$ are poorly known.

Yamamoto and Tanaka found a flow-induced change of layer spacing in a
$C_{12}E_5$ lamellar system only a few degrees away from the
equilibrium lamellar-to-sponge transition.
This does not, of course, necessarily validate our mechanism, since
any mechanism that predicts the generation of new layers will be more
susceptible in the presence of defects. In the vicinity of this
temperature shear thinning behaviour was also observed, indicating
that flow-induced change of structure was indeed occurring.  They did
not, unfortunately, report the quantitative dependence of layer
spacing on strain rate, so it is difficult to make a precise
comparison. Shear thinning has been seen in other layered systems (for
example \cite{MMK96}) but as far as we know has not been correlated
with a change of layer spacing.
  
The same $C_{12}E_{5}$ system has also been observed to form onions
under shear in other regions of the phase diagram \cite{tanakaunpub}.
The dependence of structure formation on concentration, strain rate,
shear history, and time would help to discern whether or not the
mechanism we propose is relevant for this system. Since flow modifies
the fluctuation spectrum and the steric repulsion, the shape of the
static structure factor should also change, which possibly explains
the rounding of the Bragg peak observed by Yamamoto and Tanaka
\cite{YamaTana95b}.

Note that, although defect generation is a necessary condition for
changing the layer number, the presence of too many defects should 
invalidate layer suppression effects. For example, significant 
porosity at a scale less than the patch length implies a lack of
integrity of the membrane. In this case, although on symmetry grounds the
long wavelength theory would be the standard two fluid smectic
hydrodynamics, the approximation of the microscopic membrane theory by
a simple Helfrich Hamiltonian would break down.  
For example, we would not expect the theory to apply to the
  defect-ridden SDS/decanol/water lamellar system achieved by low
  concentrations of cosurfactant, that was observed to undergo
  transitions between $a$ and $c$ orientations \cite{Berg+98b}. 
In addition there have been suggestions that in
some lyotropic lamellar systems the defect density of highly
porous systems (before any discontinuous transition) can change
continuously with the flow \cite{MinewakiKYI99}.

\subsubsection{Instability}\label{sec:instability}
The alternative scenario is that in the absence of permeation, or if
permeation is slow enough, the layer spacing cannot change from $d_0$
to $d$.  Instead however, flow would induce an effective dilational
strain $\gamma$, given by
\begin{equation}
  \label{eq:21}
  \gamma=\frac{d_0-d}{d}.
\end{equation}
At sufficiently high flow, and hence strain, the system would become
unstable to an analogue of the familiar Helfrich-Hurault effect, in
which a smectic-A liquid crystal exchanges dilational energy for
buckling energy for a large enough dilational strain. In that case,
minimisation of the smectic elastic free energy determines the
critical strain $\gamma^*$ at the onset of the instability and the
undulation wavevector $q^*_{\perp}$ \cite{DelayeRD73,pgdg}.

The analogous effect for the sheared smectic is that fluctuations are
suppressed by the flow, leading to a smaller ``preferred'' layer
spacing; in the absence of layer-creation the effect is an induced
strain, without actually applying a dilational pressure.  We can gain
a heuristic understanding of the effect by performing an equilibrium
calculation for the critical strain, and compare it to the induced
strain. The free energy of the smectic-A (Eq.~\ref{eq:5}) is adapted
\cite{pgdg} to include the non-linearity required by rotational
invariance\footnote{The correct form of Eq.~(\ref{eq:35}) involves
  $\tfrac12 (\boldsymbol{\nabla} u)^2$ rather than $\tfrac12
  (\boldsymbol{\nabla}_{\perp}u)^2$. This shifts $\gamma^*$ and $q^*$
  by higher orders of $\lambda/L$.}, as well as a long wavelength
symmetry breaking anisotropic tension arising from coarse graining the
free energy (Eq.~\ref{eq:12a}) up to the patch length, $L_p$:
\begin{equation}
\begin{split}
  \label{eq:35}
  F &=\tfrac{1}{2}\int d^3\!r \left[\bar{B}{\left[{\partial}_{z}u
        - \tfrac12{(\boldsymbol{\nabla}_{\perp}u)}^2\right]}^2 \right.  \\
  & \qquad\qquad \left. + K{({\nabla_{\perp}^2}u)}^2
    + {{\sigma\over d}(\boldsymbol{\nabla}_{x}u)^2}\right] . \\
\end{split}
\end{equation}
In contrast to a positive tension acting on the ripples, imposing the
constraints of fixed total and projected membrane areas, as in
Ref.~\cite{zilman99}, leads to a negative tension (or lateral
compression) that causes the membrane to buckle under sufficient shear
(see Appendix A).

To calculate the stability, we consider a small 
perturbation $\delta u$ around a strain induced displacement $\gamma$,
\begin{equation}
  \label{eq:36}
  u=\gamma z +\delta u\,\sin {q_z} z\cos {q_x} x \cos {q_y} y.
\end{equation}
Upon linearising in $\delta u$, we find an undulatory instability 
with an  undulation wavevector given by
\begin{align}
  \label{eq:37}
  q^*_x=0, & \qquad q^*_y=\sqrt{\frac{\pi}{\lambda L}}, & \qquad
  q^*_z=\frac\pi L, 
\end{align}
and a critical strain given by 
\begin{align}
\label{eq:26}
  \gamma^*\equiv \frac{d_0}{d^{\ast}} -1 = {2\lambda\pi\over L},
\end{align}
where $\lambda = \sqrt{K/\bar B}$ is the penetration length and $L$ is
the sample thickness. If we ignore any renormalisation of $\bar{B}$
due to the change in undulation spectrum, then the critical tension
$\sigma_c$ can be found, once $\gamma^{\ast}$ is known, from
Fig.~\ref{fig:1}.  However, the suppression of fluctuations and the
reduction in collisions diminishes $\bar B$ (see Appendix
\ref{sec:Bbar}), which increases the critical strain and hence also
increases the critical tension. An applied tension renormalised
$\bar{B}$ according to (Appendix \ref{sec:Bbar})
\begin{align}
  \bar B&={9{\pi}^2\over 64}{(k_B T)^2\over \kappa d^3}-
  {k_{B} T \sigma \over 4\pi\kappa d}     \label{eq:38}\\
  & -{{1\over 16 {\pi^2}}\left(\frac{5\pi}{4} -
      \frac{5}{\pi^2}-\frac{9}{16}\right)}{d\over\kappa}\sigma^2
  +\ldots \nonumber
\end{align}

If the strain is in the linear regime, as described by
Eq.~(\ref{eq:12f}), assuming no permeation and $a\ll d_0 \ll L$ the
effect on the reduction of $\bar B$ is small and an 
instability is induced for large enough tension given by (from
Eqs.~\ref{eq:12f}, \ref{eq:21}, \ref{eq:37}, and \ref{eq:38})
\begin{equation}
  \label{eq:51}
\sigma>\sigma_c\simeq \frac{128\pi^2 \kappa^2 \phi}{
  tk_BT L}\left(1-\frac{t k_BT}{4\pi\kappa \phi d_0}\right).
\end{equation}
If $\phi=0.2$, $k_BT=4\cdot 10^{-21}\,\text{J}$, $t\sim 3\,\text{nm}$,
$L\sim 1\,\text{mm}$ and letting $\kappa$ range from $0.2-4k_BT$ then
$\sigma_c$ is of the order $10^{-8}-10^{-5}\,\text{Jm}^{-2}$.  In
scaled parameters, this corresponds to
$\hat{\sigma}\sim10^{-3}-10^{-1}$, and a very small critical strain is
necessary, corresponding to the linear regime of Fig.~\ref{fig:1}.  If
$\sigma\sim\eta\dot\gamma d_0$ (owing to the difficulty of estimating
the prefactors in Eq.~\ref{eq:8}) and $\eta\sim 1\,\text{mPa-s}$ then
$\dot\gamma_c\sim 10^3-10^5\,\text{s}^{-1}$ for same range of
parameters.  To achieve a value comparable with experiments, Zilman
and Granek \cite{zilman99} rectified a similar discrepancy for the
critical strain rate by replacing the solvent viscosity by the
measured viscosity of the lamellar phase so that $\dot\gamma_c \sim
1\,\text{s}^{-1}$ for low $\kappa$. On the other hand, reasonable
experimental shear rates are obtained for small concentrations
($\phi\sim0.01$).  As in previous works
\cite{OswaldK82,diat93,zilman99, WunenburgerCCR00}, we have taken $L$
to be the system size. Another possibility is that $L$ corresponds to
the lamellar grain size, whose scale is set by the defect density. In
this case the reduction in $\bar B$ and hence the change in critical
strain and the strain rate will be significant.

The observation that the layer spacing does not change prior to onion
formation does not seem to have been confirmed in experiment. However,
the fact that most of the lamellar phases that undergo the onion
transition seem to exhibit Newtonian or only weakly shear thinning
flow (which is probably due to improvement in orientation of the
layers or a reduction in defect density) indicates that flow is not
bringing about such a structural change.

As might be expected, the membrane is most susceptible to undulations
transverse to the applied flow direction. Hence, this instability is
first to a stripe-like undulation, with stripes parallel to the flow
direction, or equivalently with the wavevector in the vorticity
direction.  On the other hand, an isotropic tension that penalises
Fourier modes in both directions favours square lattice buckling in
preference to stripes.  Although, as for an envisaged square lattice
buckling, the mechanism from the stripe state to further transitions
such as onions is unclear.  Our result is consistent with recent work
(unpublished) by Tanaka \textit{et al.} that shows such an undulation
in a $C_{12}E_5$ surfactant lamellar phase \cite{tanakaunpub}.
Similar behaviour has recently been demonstrated with simulations of a
thermotropic lamellar phase \cite{Soddemann}.  The latter does not
correspond to a Helfrich-stabilised lamellar system, while the
experiments probably do.  Recent experiments on $C_{10}E_5$ by Zipfel
\textit{et al.}  revealed a kinetic intermediate between the
$c$-oriented lamellar and onion phases, compatible with cylindrical
multilamellar ``leeks'' \cite{ZipfelNLLOR01}. Such leeks are
compatible with an initial undulation in the layers, with wavevector
parallel to the vorticity direction, that subsequently breaks the
layers and stabilises in the cylindrical symmetry. We also note that
the transient transitions to leeks and then onions occurred at roughly
the same strain (or order a few thousand strain units), for different
strain rates.

We emphasise that we have ignored Gaussian curvature, which is 
likely to influence transitions involving layer topology changes under shear.
In equilibrium, reduction of salinity \cite{leon00,Lekkerk90} 
or temperature \cite{LeOM00} or cosurfactant
\cite{PorteABM89,*SkouriMAP91} are all known to decrease the Gaussian
curvature modulus $\bar\kappa$, thus favouring phases consisting of
spherical over bicontinuous structures.  A $C_{10} E_3$/water solution
developed an onion transition upon reduction of the temperature at an
imposed shear rate \cite{LeOMZR01}, which is consistent with the known
decrease of $\bar\kappa$ with decreasing temperature in this system.
Similarly, a low salt ionic surfactant AOT lamellar system showed no
critical strain rate \cite{leon00,ZipfelNLLOR01}, but rather onion
formation at a characteristic critical applied strain with a time that
increased dramatically for higher salt concentration and thus larger
$\bar\kappa$. In this particular case, onion formation was independent
of strain rate, suggesting a mechanism different from that we have
proposed.  Finally, at high enough concentrations of cosurfactant, the
SDS/decanol/water system seemed to be prevented from forming onions
\cite{ZipfelBLR99}, even though the system was less permeable. Here we
suggest the increase in Gaussian curvature may also be correlated with
this observation.

\section{Summary}\label{sec:conclusion}
We have studied the suppression of the undulations in a
Helfrich-stabilised lyotropic lamellar phase in shear flow, by crudely
modelling the flow as an effective anisotropic tension. This decreases
the intermembrane compression modulus $\bar{B}$ due to a reduction in
fluctuations, and correspondingly changes the structure factor. There
are two general consequences, depending on the permeability of the
lamellar phase (under shear).

\begin{enumerate}
\item If permeation or defects allow the generation of new layers,
  such a system would eventually attain a new layer spacing,
  consistent with the flow-induced reduction of collisions.
\item If the system cannot change the number of layers, or the process
  is very slow, then either:
\begin{enumerate}
\item The system can maintain the original layer spacing for low
  strain rates, with a concomitant effective strain due to the
  deviation from the preferred stable steady state.
\item For higher strain rates, an instability to undulations along the
  vorticity direction can relieve the effective induced strain. A
  linear analysis does not allow us to determine whether or not such a
  state would result in a stable undulatory phase (as reported in
  simulations \cite{Soddemann} and experiments
  \cite{tanakaunpub}), become unstable to ripping layers to produce either
  cylinders with axes parallel to the velocity direction, as seen in a
  number of systems, or produce onions. In such a situation the layer
  spacing prior to forming the new state (striped undulation,
  cylinders, or onions) would not be expected to show a change in
  layer spacing.
\end{enumerate}
\end{enumerate}

Although the reduction of $\bar B$ has a very small effect on the 
critical strain (or equivalently, the critical applied tension or
strain rate) at which an undulatory instability may occur, the
reduction in repulsion allows any attractive potential to become more
significant. For highly swollen lamellar phases close to an
equilibrium unbinding transition \cite{milnerroux92} there is the
possibility of a flow-induced unbinding, which would probably be
manifested in solvent expulsion or macroscopic phase separation.

In addition, we have discussed possible sources for the tension,
including bilayer collisions, intra-bilayer dissipation and
convection. The latter mechanism leads to a scale-dependent
``tension'' upon coarse-graining from the microscopic layer position 
$h(x,y)$ to the mean smectic layer displacement $u(x,y,z)$. 

One implication of our result is that, in principle, lamellae and
onions in flow are generally expected to have different layer spacings
and hence different concentrations.  Onions, particularly the large
ones that appear at low shear rates and are the likely candidates for
coexistence with lamellae \cite{diat93}, are very close in free energy
and hence layer spacing to the equilibrium lamellar phase, while the
lamellae subject to shear flow could be subject to strong modification
of their fluctuation spectrum. This would imply slow kinetics for the
eventual transformation to onions, due to concentration changes, and
non-trivial shapes for the measured flow curves, or `plateaus', in
regions of macroscopic coexistence \cite{olmsted99b}.

Our work has focused on an instability to onion formation. We have
not determined whether or not onions are metastable above a given
strain rate, in which case the instability would be the analogue of an
equilibrium spinodal. Indeed, it would be very interesting to study
experimentally the history- and time-dependence upon cycling the
strain rate or shear stress (both increasing and decreasing) through
the lamellar-onion transition. Experiments on
SDS/decanol/dodecane/water have indicated regions of concentration,
or layer spacing and degree of crumpling, in which the transition is
apparently continuous (low concentration or larger equilibrium layer
spacing) and first order (high concentration and smaller layer
spacing) \cite{DRN93,RND93}.  In addition, the history dependence may
alter the eventual point of instability. For example, if the strain
rate is increased slowly compared to the time scale associated with
permeation, a reduction in layer spacing may be observed which would
yield a higher critical strain rate.  Further, if a system displays a
shear induced reduction in layer spacing, onions might be formed if a
strain rate larger than the critical strain rate is applied faster
than permeation effects can occur.

We have assumed that flow acts as an effective tension. This is
obviously quite crude, and forthcoming work will study the dynamics of
individual membranes in flow. For membranes in the
  $a$-orientation, as studied by Ramaswamy to describe a layer
  collapse transition \cite{Ramaswamy92b,alkahwaji00}, the convective
  term is linear, while in the $c$ orientation, the convective term is
  non-linear, as in Eq.~(\ref{eq:2}). On symmetry grounds, upon
coarse-graining the fluctuations to a scale of order the patch size
$L_p$ between collisions, the convective non-linearity generates a
tension-like restoring term in the long-wavelength dynamics,
Eq.~(\ref{eq:13}), which depends on the single membrane
  relaxation mechanism and is also, strictly, dependent on length
  scale and wavenumber in a very different form than a conventional
  ``tension''.  
\begin{acknowledgement}
  We thank H Tanaka, T Kato, J Penfold, D Roux, AJ Bray, TCB McLeish,
  D Bonn, J Leng, and K Kremer for helpful discussions.
\end{acknowledgement}
\appendix
\section{Relation to Previous Work} \label{sec:ZG work}
In a related work, Zilman and Granek (ZG) \cite{zilman99}, studied the
effect of shear flow on the $c$-orientation of lamellar surfactant
systems, to address the lamellar-to-onion transition. Our treatment
differs from theirs in several respects. ZG advocate the physical
picture that, for cylindrical Couette flow, layers retain their 
integrity so that the projected area is fixed by the experimental
geometry. Shear flow removes small scale fluctuations and hence would
stretch the projected area of a free membrane. If this projected area
is constrained, then the membrane can only undergo macroscopic
buckling to redistribute undulations from small scale crumpling to
larger scale undulations. Hence ZG model the coarse-grained membrane
as experiencing a \textit{negative} buckling tension. Our approach is
quite different, and consists in examining the effect of a positive
tension-like quantity on the underlying crumpling spectrum and hence
the steady state layer spacing.

The ZG picture is certainly relevant if for example, most layers in a
cylindrical Couette geometry have cylindrical topology with edges only
at the cylinder ends. On the other hand, if most layers in the system
do not encircle the centre Couette cylinder but end in edges, then
they need not maintain a given projected area, but can adjust by
moving defects. We feel that this relaxed constraint applies more
generally than the fixed projected area constraint, but we acknowledge
that line dislocation densities are notoriously difficult to determine
experimentally.
 
Before outlining other differences we review the different layer
variables for a lyotropic smectic:
\begin{subequations}
\begin{align}
  h(x,y) & =
  \begin{cases}
&\textrm{microscopic layer position}\\
  &(a < dx,dy< L)\\
  \end{cases}\\
u(x,y,z) & =
  \begin{cases}
    &\textrm{smectic layer displacement}\\
    &(L_p < dx,dy,dz < L),
  \end{cases}
\end{align}
\end{subequations}
where $L$ is the system size, and $dx, dy,$ and $dz$ refer to
differences of the independent variables $x,y,z$. The broken symmetry
variable $u(x,y,z)$ describes the average layer displacement.  In the
undeformed state a layer is taken to be ``flat'', $u(x,y,z)=0$.  In
principle, $u(x,y,z)$ can be obtained by coarse-graining the highly
crumpled microscopic variable $h(x,y)$ up to a length scale roughly of
order the collision length $L_p$; hence $u(x,y,z)$ is defined only on
length scales larger than this cutoff.
Conversely, $h(x,y)$ is defined
down to a microscopic scale $a$, of order a surfactant head diameter.
In performing the coarse-graining information about smaller length
scales is retained, and resides in the other hydrodynamic variable,
the concentration, which essentially measures the degree of crumpling
at length scales smaller than the collision length $L_p$. Hence the
free energy for lyotropic smectics is most naturally defined in terms
of average layer displacement and concentration changes, although
other combinations of these basic degrees of freedom are, of course,
possible.

ZG draw an important distinction between different ``projected
areas''. ZG define the geometric area $A_{geom}$ as the conventional
projected area $A_\perp$ imposed by sample geometry. For parallel
plate rheometer $A_\perp$ is the plate area, while for a Couette
rheometer of cylinder height $h$, $A_\perp$ is the cylindrical area
$2\pi R h$ at a given radius $R$.  ZG distinguish this area from a
``physical'' projected area $A_{phys}$, which can be interpreted as
the area of the coarse-grained smectic variable $u$; \textit{i.e.} (in
the Monge gauge)
\begin{equation}
  \label{eq:10}
  A_{phys} = \int_{A_{\perp}}d^2r\left[1 + 
  \tfrac12\left(\boldsymbol{\nabla}_{\perp}
      u\right)^2\right].
\end{equation}
This corresponds to the ``constraint''
introduced in Eq.~(ZG-19), which is in fact not a constraint but a
definition, as follows by examining their Figure~3 in conjunction with
their discussion introducing physical projected area ($A_o$ in their
notation, in Figure 3 of their work).

By modifying the derivation of the Helfrich interaction ZG derive an
expression for the macroscopic free energy $F_{ZG}[u,\Delta]$ (Eq.
ZG-14a) of a lyotropic smectic as a function of the mean layer
displacement variable $u(x,y,z)$ and the local change in physical
projected area,
\begin{equation}
  \label{eq:15}
  \Delta = \frac{\delta A_{phys}}{A_{phys}^{(0)}}.
\end{equation}
However, $\Delta$ cannot be an independent local hydrodynamic
variable, like concentration (or volume fraction) in the standard
two-fluid smectic description, because it is rigorously defined
through Eq.~(\ref{eq:10}) as a non-local function of $u(\vec{r})$.
Hence, $\Delta$ does not contain, \textit{a priori}, any information
about small scale crumpling or concentration.

Although ZG claim that $\Delta$ may be eliminated in favour of the
local volume fraction $\phi$, and then ``integrated out'' to recover
the free energy as a function of $u(\vec{r})$ at constant chemical
potential, for which layer compressions are penalised by the usual
modulus $\bar{B}$ (ZG:Appendix B), this is not correct.
The local volume fraction is defined by (Eqs.~\ref{eq:9} and ZG-B.1)
\begin{equation}
  \label{eq:16}
  \phi = \frac {t A}{d A_{\perp}}.
\end{equation}
ZG use this relation, at fixed membrane area $A$ rather than fixed
projected area $A_{\perp}$ (Eq.~ZG-B2), to relate changes in layer
spacing, concentration, and $A_{\perp}$:
\begin{equation}
  \label{eq:17}
  \frac{\delta\phi}{\phi} + \frac{\delta d}{d} = - \frac{\delta
    A_{\perp}}{A_{\perp}}.
\end{equation}
They assert that $A_{\perp}$ may be taken to be the physical projected
area $A_{phys}$ and apply Eq.~(\ref{eq:15}) to relate $\delta\phi$, 
$\delta d$ and $\Delta$, but this is inconsistent with the geometric
definition above (Eqs.~\ref{eq:10} and ZG-19). They have
already defined $\delta d$ in terms of $u$, using 
the geometric relation (ZG-13)
\begin{equation}
  \label{eq:18}
  \frac{\delta d}{d} = \frac{\partial u}{\partial z} - \tfrac12
  \left(\boldsymbol{\nabla}_{\perp} u\right)^2,
\end{equation}
correct to order $\theta^2$ for small rotations $\theta$.
Combining Eqs.~(\ref{eq:10}, \ref{eq:17} and \ref{eq:18}) gives
\begin{equation}
  \label{eq:22b}
\frac{\delta\phi}{\phi}=-\frac{\partial u}{\partial z} . 
\end{equation}
Thus a small rigid body rotation\footnote{In general,
$u=-x\sin\theta+z(1-\cos\theta)$.} about the $\hat{\vec{y}}$ axis,
$u\backsimeq-x\theta+\tfrac12 z\theta^2$ leads to the unphysical result
$\delta \phi/\phi =-\tfrac12 \theta^2$.

Similarly, in the absence of the buckling tension $\sigma$ (which they
argue vanishes for zero shear) the final free energy ZG use (Eq.
ZG-21a), obtained from Eq.~(ZG-14a) by implementing the definition of
physical area (Eq.~\ref{eq:10}), is not invariant under small uniform
rotations for $\sigma=0$.  ZG also consider a scenario in which the
membrane area can vary at fixed layer number. In this case they
minimise $F_{ZG}[u,\Delta]$ over $\Delta$ and use the resulting free
energy to determine stability; however, this is inconsistent with the
definition of $\Delta$, according to Eq.~(\ref{eq:10}).

\onecolumn
A final problem concerns the non-linear analysis. The correct form of
the rotational invariant term in the free energy (Eq.~\ref{eq:35} and
ZG-14a) involves $\tfrac12 (\boldsymbol{\nabla} u)^2$ rather than
$\tfrac12 (\boldsymbol{\nabla}_{\perp}u)^2$. The effect of this
approximation on the linear stability is negligible; the critical
strain and wavevector shift by order $\lambda/L$ where $\lambda$ is
the penetration length.  However, a substantial correction is expected
for a non-linear analysis of the buckled state, as in ZG-Section 4.2.
\section{Calculation of Tension-Renormalised Compression
  Modulus $\bar B(\sigma)$} \label{sec:Bbar}
To compute $\bar B$ we follow the mean field variational approach of
Rabin and Bruinsma \cite{bruinsmarabin92}, who calculated the free
energy difference between a confined system and a corresponding set of
free layers, and add an anisotropic tension. A related calculation has
been performed by Lubensky \textit{et al.} \cite{lubensky90}. First,
we replace the free energy of interacting membranes by an equivalent
single membrane,
\begin{equation}
  \label{eq:40}
 f ={1\over 2}\int d^2r \left[\kappa{(\nabla^2_\perp h)}^2+
    (\sigma+\mu){(\boldsymbol{\nabla}_xh)}^2 +\Gamma{h^2} \right],
\end{equation}
where $\Gamma$ and $\mu$ are variational parameters that maintain the
equilibrium layer spacing, while $\sigma$ penalises changes in excess
area due to undulations in the $x$-direction. Assuming constant
membrane thickness and area per molecule, $\mu$ is an effective
chemical potential. For convenience, we consider a reservoir that
admits material along the $x$ direction and also penalises excess
area.  The last term is a harmonic potential that mimics membrane
interactions, and thus determines the mean layer spacing.  

The equilibrium distribution of membrane height fluctuations is given
by
\begin{equation}
  \label{eq:41}
  \rho_\perp(h)={\exp\nts\left[{-{1\over
  2}\beta \sum_{\vqp}\nts(\mubsiguni)\left|h_{\vqp}\right|^2}\right]\over N},
\end{equation}
where $\beta=1/k_B T$ and
\begin{equation}
  \label{eq:42}
  N=\int\nts\prod_{\vqp}dh_{\vqp}\nts\exp\nts\left[{-{1\over
  2}\beta\sum_{\vqp}\nts(\mubsiguni)\left|h_{\vqp}\right|^2}\nts\right].
\end{equation}
By the equipartition theorem the mean layer fluctuations are 
\begin{align}
  \label{eq:43}
  \langle h^2(\vec{r})\rangle
  = {k_B T\over {4\pi^2}}\int\frac{d^2\vqp}{\mubsiguni}.     
\end{align}

The relation of $\langle h^2\rangle$ to layer spacing
(Eq.~\ref{eq:7e}) determines $\Gamma$ as a function of $\mu$, while
$\mu$ is finally determined by minimising the mean membrane free
energy, given by (using Eq.~\ref{eq:41})
\begin{subequations}  
\begin{align}
  F&=\langle f \rangle + k_{B}T\langle\ln\rho_\perp\rangle.
  \label{eq:44}\\
  &=\sum_{\vqp}\left[{1\over 2}\left(\kappa q_{\perp}^4+
    \sigma q_x^2 \right)\langle\left|h_{\vqp}\right|^2\rangle
  +k_BT\left\langle\ln{
\left({\exp{-{1\over2}\beta\left[\mubsiguni\right]
      \left|h_{\vqp}\right|^2}\over N} 
    \right)}\right\rangle\right].  \label{eq:45}
\end{align}
\end{subequations} 
The free energy difference between free and confined membranes is
\begin{equation}
  \Delta F
  ={1\over 2} k_BT\sum_{\vqp}\frac{-(\mu q_{x}^2+\Gamma)}{\mubsiguni} 
   +\ln\left[\frac{\mubsiguni}{\kappa q_{\perp}^4 +\sigma q_{x}^2} \right].
  \label{eq:46}  
\end{equation}

Expressing $\Gamma$ to second order in $(\mu+\sigma)$,
\begin{equation}
  \label{eq:47}
\Gamma(\mu+\sigma)=\Gamma_0+\Gamma'_0(\mu+\sigma)
+{1\over 2}\Gamma''_0(\mu+\sigma)^2 +..., 
\end{equation}
we find, using Eqs.~(\ref{eq:7e}, \ref{eq:43}),
\begin{align}
  \label{eq:47b}
  {\Gamma}_0 &= \frac{1}{\beta^2\kappa {(8\alpha d^2)}^2}\,,&
  {\Gamma '}_0 &= -{1\over {8\pi\alpha\beta^2\kappa d^2}}\,,&
  {\Gamma ''}_0 &= {1\over \beta^2\kappa}\left[{3\over 16}- {27 \over
    16 {\pi}^2} \right].
\end{align}
For $\sigma =0$, $\Gamma_0$ is the same while ${\Gamma '}_0 =
-1/4\pi\alpha\beta^2\kappa d^2$ (correcting a sign error in
\cite{bruinsmarabin92}) and ${\Gamma ''}_0 = (\tfrac12 -
\tfrac{2}{\pi^2})\beta^2\kappa$. Substituting Eqs.~(\ref{eq:47}) and
(\ref{eq:47b}) into Eq.~(\ref{eq:46}) gives, after some calculation,
\begin{equation}
{\Delta F\over A_\perp} = {1\over 128}{{1\over\kappa{\beta}^2 \alpha
    d^2}} + {1\over 8}{\left({27\over
      16}-{1\over\pi^2}\right)}{{\alpha
    d^2\over\kappa}{(\mu +\sigma)^2}} 
+{\sigma\over 16\pi\beta\kappa}\left[\ln{{\sigma^2\over
      {\kappa\Gamma}}}-4\ln{2} -{5\over 2}\pi^2\alpha d^2(\mu+\sigma)\right]. 
\label{eq:48}
\end{equation}
The compression modulus at constant chemical potential is given by
\begin{equation}
  \label{eq:49}
   \bar B = d \left({{{\partial}^2}\over{\partial d^2}}\right)
\left({{\overline{\Delta F}}\over {A_\perp}}\right),
\end{equation}
where $\overline{\Delta F}$ is the minimum free energy with respect
to the chemical potential, yielding
\begin{equation}
   \label{eq:50}
  \bar B={9{\pi}^2\over 64}{(k_B T)^2\over \kappa d^3}- 
  {k_{B} T \sigma \over 4\pi\kappa d}  
  -{{1\over 16 {\pi^2}}\left(\frac{5\pi}{4} -
      \frac{5}{\pi^2}-\frac{9}{16}\right)}{d\over\kappa}\sigma^2 +\ldots
\end{equation}
As expected, Seifert's calculation of the potential between a membrane
and vesicle in the presence of a small tension reduces to a similar
result \cite{Seif95}
\begin{equation}
  \label{eq:11}
  \bar B ={9{\pi}^2\over 64}{(k_B T)^2\over \kappa d^3}-
  {\pi k_{B} T \sigma \over 32 \kappa d} +\ldots 
\end{equation}
\twocolumn
\bibliography{membs,articles,onions,helfrich,rheofolks,books,newbib,ryan,cit2}
\bibliographystyle{epjsty}
\end{document}